\documentclass[dvips]{article}
\usepackage{icrctc07}

\title{Differences in dE/dX for $\mu+$ and $\mu-$ \\
and its Effect on the Underground Charge Ratio}
\shorttitle{dE/dX for $\mu+$ and $\mu-$ and the Underground Charge Ratio}

\authors{Juergen Reichenbacher$^{1}$,\, {\em Presenter}: Maury Goodman$^{1}$ }
\shortauthors{Juergen Reichenbacher}
\afiliations{$^1$Argonne National Laboratory,\, USA}
\email{contact: Juergen.Reichenbacher@anl.gov}

\abstract{Theoretical calculations predict a small fractional difference in 
energy loss for $\mu^+$ and $\mu^-$ of the order of $0.15\,\%$ at high 
energies. This is predominantly due to a $z^3$ term in an extended ionization 
$dE/dX$ relation, in analogy to the Barkas effect at low energies 
around the Bethe-Bloch maximum. The atmospheric muon energy spectrum is 
steeply falling off with approximately $E^{-3.7}$ and thus the small 
difference in $dE/dX$ between $\mu^+$ and $\mu^-$ 
at high energies results in 
an amplified charge asymmetry of about $0.6\,\%$ a few thousand meters water 
equivalent deep underground.}

\begin{document}
\maketitle

\section{Introduction}
The MINOS far detector
is the first large underground experiment with a magnet that can 
measure the ratio of $\mu^+$ to $\mu^-$ with high 
precision.\cite{bib:mupaperFNAL,
bib:mufson}.   A precise measurement of the charge ratio can then
be used to ascertain special properties of the cosmic ray showers,
such as the $\pi^+/\pi^-$ ratio and the $K^+/K^-$ ratio\cite{bib:interpret}.
The survival probability for muons to reach an underground
detector depends on the energy loss, so if there is any difference in 
energy loss between $\mu^+$ and $\mu^-$, that would affect the measured
charge ratio.  The statistical error on the MINOS measurement is
remarkably small, and MINOS reports:
\begin{equation}
r = \frac{N(\mu^+)}{N(\mu^-)} = 1.374 \pm 0.003(stat)^{+0.012}_{-0.010}(sys)
\end{equation}
for surface muons with an energy near 1 TeV or higher.
Thus, even very small differences in the energy loss between  $\mu^+$ 
and $\mu^-$ could be important in the interpretation of these measurements.
\par
The statistical energy loss of muons, traversing an amount $X$ of matter in 
$g/cm^2$, with energies far above the Bethe-Bloch minimum is usually 
parameterized as
\begin{equation}
- \frac{dE_{\mu}}{dX} = a(E_{\mu}) \,+\, 
\displaystyle\sum_{n=1}^{3} b_{n}(E_{\mu}) \cdot E_{\mu},
\label{eq:2}
\end{equation}
where $a$ is the collisional term (i.e. ionization,
mostly due to delta-ray 
production) and $b$ in the second term accounts for the three radiative muon 
energy loss processes: 1.\,Bremsstrahlung and 2.\,pair production,
as well as 
3.\,photonuclear interactions. In {\em Table} \ref{table1} these energy loss 
parameters are listed for standard rock. The critical energy 
where ionization 
losses equal radiative losses in standard rock is approximately $0.6\,$TeV. 
The average muon energy for a muon which reaches the depth of MINOS is 
greater than $1$\,TeV, so the $b$ term and its energy dependence are important 
in calculating the energy loss. This paper focuses on the (small) differences 
in the $a$ and $b$ terms for $\mu^+$ and $\mu^-$. 

\begin{table}[h]
{\tiny 
\begin{tabular}{|c||c||c|c|c|c|}
\hline
$E_{\mu}$ & $a_{ion}$ & $b_{brems}$ & $b_{pair}$ & $b_{DIS}$ & $\Sigma b$ \\
\cline{3-6}
 [$GeV$] & [$MeV\,cm^{2}/g$] &\multicolumn{4}{c|}{[$10^{-6}\,cm^{2}/g$]} \\
\hline\hline
$10$ & 2.17 & 0.70 & 0.70 & 0.50 & 1.90 \\
$10^{2}$ & 2.44 & 1.10 & 1.53 & 0.41 & 3.04 \\
$10^{3}$ & 2.68 & 1.44 & 2.07 & 0.41 & 3.92 \\
$10^{4}$ & 2.93 & 1.62 & 2.27 & 0.46 & 4.35 \\
\hline
\end{tabular}
}
\caption{{\small Average muon energy loss parameters calculated for 
standard rock \cite{bib:Groom}\cite{bib:pdg}}}\label{table1}
\end{table}

\section{Calculated Difference in Ionization dE/dX for $\mu^+$ and $\mu^-$}

\par
\begin{figure}
\begin{center}
\noindent
\includegraphics [width=0.5\textwidth]{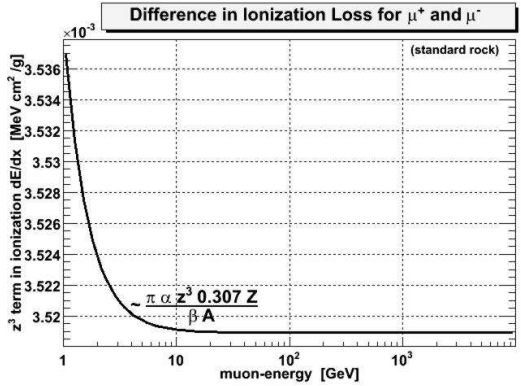}
\end{center}
\caption{Calculated difference in ionization energy loss between
positive and negative muons in standard rock (average nuclear properties:
$\overline{Z}=11$, $\overline{A} = 22$ \,\cite{bib:pdg}).}
\label{fig:fig1}
\end{figure}

 At low energies, around the Bethe-Bloch maximum, the difference in 
ionization energy loss is known as the Barkas effect\cite{bib:fermi}, 
and there have been efforts to both measure and calculate 
those differences\cite{bib:mccarthy}. 
Calculations show that negative particles lose energy at a slower 
rate, with the difference dropping from tens of percent at MeV energies 
to about $0.3$\,\% in the GeV range. 
Such differences were experimentally verified both at MeV 
energies\cite{bib:barkas1}-\cite{bib:barkas3} 
and in the GeV range \cite{bib:barkas4}. 
At higher energies, 
this difference in ionization energy loss has usually been neglected, 
and we are not aware of any measurements. As described in Reference 
\cite{bib:jackson}, the usual ionization 
energy loss term for muons (of either sign) 
depends on $z^2$, and the difference between $\mu^+$ and $\mu^-$ arises from 
a small additional $z^3$ correction term. 
This correction term in $dE/dx$ is:
\begin{equation} 
\displaystyle\left(\frac{dE}{dX}\right)^{corr}_{ion} = \frac{\pi \alpha z^3 0.307 Z}{2 \beta A} \quad 
[MeV\, cm^{2}\, g^{-1}]
\label{eq:3}
\end{equation}
where $\alpha$ is the fine structure constant, $z$ is the charge, $\beta$ is 
the relativistic velocity, and $Z$ and $A$ are the nuclear properties of the 
material through which the muon is passing. The absolute value of the 
difference in ionization energy loss between positive and negative muons in 
standard rock\cite{bib:pdg} is plotted in Figure \ref{fig:fig1}. 
It is fairly constant above $10$\,GeV, at a value corresponding to 
approximately $0.15$\,\% of the mean energy loss in the ionization dominated 
energy regime ({\em c.f.} Table \ref{table1}). 

\section{Calculated Difference in Bremsstrahlung dE/dX 
for $\mu^+$ and $\mu^-$}

\begin{figure}
\begin{center}
\noindent
\includegraphics [width=0.5\textwidth]{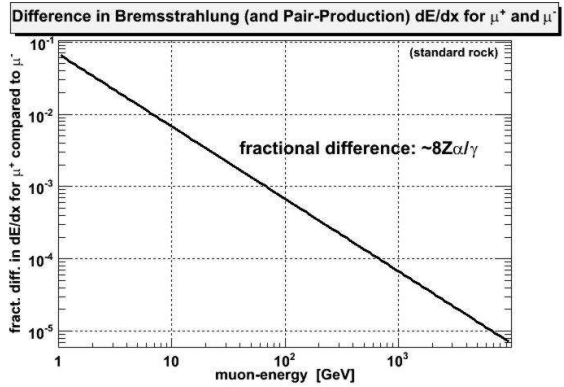}
\end{center}
\caption{Calculated fractional difference in Bremsstrahlung energy loss 
between positive and negative muons.}
\label{fig:fig2}
\end{figure}

Above an energy near $0.6$\,TeV in standard rock, radiative energy loss 
becomes comparable
to ionization energy loss, and continues to grow at
higher muon energies. 
From Reference \cite{bib:jackson2}, the fractional difference in 
Bremsstrahlung energy loss between positive and negative muons is 
\begin{equation}
\frac{[\frac{dE}{dX}]^{\mu+}_{brems} - [\frac{dE}{dX}]^{\mu-}_{brems}}{[\overline{\frac{dE}{dX}}]_{brems}} = \frac{8 Z \alpha}{\gamma}
\label{eq:4}
\end{equation}
where $\gamma$ is the Lorentz factor of the muon. 
Again, the $\mu^+$ has a slightly higher energy loss. 
This fractional difference {\it decreases} with energy and is already 
negligible where radiative energy losses become important. 
This fractional difference is plotted in Figure \ref{fig:fig2} for muons in 
standard rock. 
Presumably, the same fractional difference can also be assigned for 
pair-production, as the underlying process is a two-photon exchange between 
the muon and the constituents of the nucleus, and thus the cross sections for 
$\mu^+$ and $\mu^-$ should scale in the same way as for Bremsstrahlung. 

\section{Muon Range Underground for $\mu^+$ and $\mu^-$}

Taking the vertical muon intensity from an optimized Gaisser parameterization 
of the muon flux at the surface and propagating this energy spectrum 
underground according to statistical ionization and radiative energy losses, 
it is possible to precisely calculate the underground muon intensity. 
This procedure is described in detail in \cite{bib:myCrouchProc} for 
overburdens of standard and Soudan rock (MINOS). 
First, the average muon range underground, for each value of surface energy, 
has to be precisely computed. For this, the energy dependent $a$ and 
$\Sigma b$ values are parameterized for standard and Soudan rock as in 
\cite{bib:myCrouchProc}. The additional ionization loss 
according to Eq. 3 is then added for $\mu^+$ to the value of the function for 
$a$ (subtracted for $\mu^-$). 
Conservatively, $90\,\%$ of the value of the function for $\Sigma b$ (the 
total radiative losses), are scaled with the energy dependent fractional 
difference $R=8Z\alpha/\gamma$ according to Eq. 4 as $\Sigma b^{\pm} = 0.9 
\cdot \Sigma b \cdot (1 \pm R/2)$ for $\mu^+$ and $\mu^-$, respectively. 
The contribution from photonuclear production (DIS), which accounts for a 
constant fraction of $10\,\%$ of all radiative muon energy losses in standard 
rock (in the region of interest from $250\,$GeV to $10\,$TeV) is not scaled 
with the fractional difference. 
According to differential Eq. 2 the propagation of the muon energy is then 
numerically computed, separately for $\mu^+$ and $\mu^-$ (in standard and 
Soudan rock, respectively). Thus, for each initial value of muon energy,
the slant depth in meter-\-water-\-equivalent where the 
muons of different charge range out is determined.

\section{Muon Charge Asymmetry from Ratio of Intensities Underground}

\begin{figure}
\begin{center}
\noindent
\includegraphics [width=0.5\textwidth]{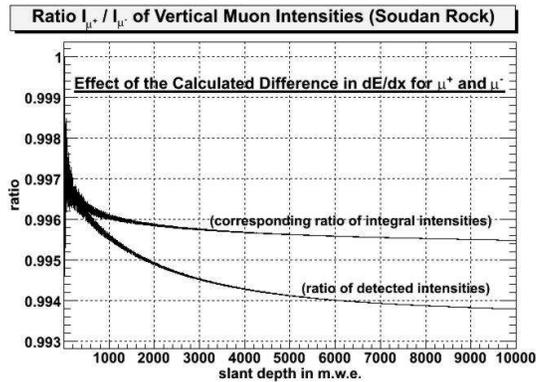}
\end{center}
\caption{Calculated ratio of positive to negative vertical muon intensities in 
Soudan rock as a function of slant depth. The upper curve is for all muons, 
the lower curve is for muons with a remnant momentum of less than 250 GeV/c 
($\approx$ the maximum detectable momentum in the MINOS far
detector).}\label{fig:intensity}
\end{figure}

Using the average muon range underground, calculated for positive and 
negative muons in rock as described in the last section, and an optimized 
Gaisser parameterization of the differential intensity of vertical 
muons at the surface, we have computed the corresponding 
underground intensities of positive and negative muons as a function of slant 
depth for a given rock composition. 
The resulting ratio of the $\mu^+$ and $\mu^-$ intensity curves is shown in 
Figure \ref{fig:intensity} for Soudan rock. The upper curve corresponds to the 
fractional difference in integral intensities of $\mu^+$ and $\mu^-$ at a 
given slant depth. For slant depth values above about $1000\,mwe$ the underground 
ratio $N(\mu^+)/N(\mu^-)$ is lowered by roughly $0.4\,\%$. However, since 
the 
charge of only the lower energy muons can be identified in a magnetic detector, 
owing to its maximum detectable momentum \cite{bib:mdm}, the detected intensity 
corresponds to the charge ratio of the muons at depth below some momentum. 
The lower curve in Figure \ref{fig:intensity} depicts the fractional 
difference in intensity for underground muon momenta below $250$\,GeV/c, 
corresponding to the approximate maximum detectable momentum of MINOS. For 
increasing slant depth values the measured underground ratio 
$N(\mu^+)/N(\mu^-)$ is further  reduced and saturates at
about $0.6\,\%$ below its surface value for slant depths larger than roughly $5000\,mwe$. 
\par The dominant $0.15$\% difference in ionization energy loss between $\mu^+$ and 
$\mu^-$ gets amplified by a factor of about $3.7$, due to the approximate 
$E^{-3.7}$ dependence of the differential muon spectrum. 
The impact of the rock composition is almost negligible, as the induced 
muon charge asymmetry under Soudan rock lowers 
the surface value of the ratio 
by an additional amount less than $0.02\,\%$ compared to standard rock. 

\section{Summary}

There is a small fractional difference in energy loss for $\mu^+$ and $\mu^-$ 
of the order of $0.15\,\%$ predicted by theoretical calculations at high 
energies, predominantly due to a $z^3$ correction term in the ionization 
energy loss. This causes that measurements of the atmospheric muon charge 
ratio $N(\mu^+)/N(\mu^-)$ deep underground (e.g. with the MINOS detector), to
observe a slightly lower ratio than at the surface. 
Moreover, as the atmospheric muon energy spectrum is steeply falling off with 
approximately $E^{-3.7}$, the small difference in energy loss between $\mu^+$ 
and $\mu^-$ at high energies results in an amplified charge asymmetry of about 
$0.6\,\%$ several thousand meters water equivalent deep underground. 
The calculations presented herein allow for a correction of the underground 
measured muon charge ratio to its surface value.

\section{Acknowledgments}

This work was supported by the U.S. Department of Energy (DOE) under contract 
DE-AC02-06CH11357. I also like to 
thank Geoff Bodwin from the High Energy Physics Division at Argonne, 
Stan Wojcicki from Stanford,
Stuart Mufson from the University of Indiana,
Don Groom from Lawrence Berkeley National Laboratory as well as 
Thomas Fields and
the neutrino physics group at Argonne for valuable discussions and in 
particular Maury Goodman for his support.


\hfil\break

\end{document}